\documentclass[sigconf]{acmart}

\usepackage{subcaption}
\usepackage{svg}

\usepackage{enumitem}
\setlist{nosep}
\usepackage{tabularx}
\usepackage{multirow}

\usepackage{xspace}
\makeatletter
\DeclareRobustCommand\onedot{\futurelet\@let@token\@onedot}
\def\@onedot{\ifx\@let@token.\else.\null\fi\xspace}

\AtBeginDocument{%
  }

\acmSubmissionID{4499}

\begin{document}

\title{GraphStory: Collaborative Story Writing through Event-Based Narrative Editing}


\author{Xuan-Vu Le}
\authornote{Both authors contributed equally to this research.}
\orcid{0009-0009-6094-4912}
\affiliation{%
  \institution{University of Science}
  \city{Ho Chi Minh}
  \country{Vietnam}
}
\affiliation{%
  \institution{Vietnam National University}
  \city{Ho Chi Minh}
  \country{Vietnam}
}

\author{Minh-Loi Nguyen}
\authornotemark[1]
\orcid{0009-0003-2630-3325}
\affiliation{%
  \institution{University of Science}
  \city{Ho Chi Minh}
  \country{Vietnam}
}
\affiliation{%
  \institution{Vietnam National University}
  \city{Ho Chi Minh}
  \country{Vietnam}
}

\author{Khanh-Duy Le}
\orcid{0000-0002-8297-5666}
\affiliation{%
  \institution{University of Science}
  \city{Ho Chi Minh}
  \country{Vietnam}
}
\affiliation{%
  \institution{Vietnam National University}
  \city{Ho Chi Minh}
  \country{Vietnam}
}

\author{Minh-Triet Tran}
\orcid{0000-0003-3046-3041}
\affiliation{%
  \institution{University of Science}
  \city{Ho Chi Minh}
  \country{Vietnam}
}
\affiliation{%
  \institution{Vietnam National University}
  \city{Ho Chi Minh}
  \country{Vietnam}
}

\author{Trung-Nghia Le}
\orcid{0000-0002-7363-2610}
\affiliation{%
  \institution{University of Science}
  \city{Ho Chi Minh}
  \country{Vietnam}
}
\affiliation{%
  \institution{Vietnam National University}
  \city{Ho Chi Minh}
  \country{Vietnam}
}
\authornote{Corresponding author. Email: ltnghia@fit.hcmus.edu.vn}


\begin{abstract}
Story writing is a popular yet complex creative activity that requires organization of ideas and iterative exploration, particularly during early-stage ideation. While many AI-based writing assistants have been developed, existing approaches primarily focus on generating long-form coherent text and improving user controllability during text production, providing limited support for brainstorming, connecting ideas, and validating alternative narrative flows. We present GraphStory, an interactive writing support system that leverages a graph-based representation to provide a comprehensive view of narrative structure and facilitate ideation. The system enables users to organize and connect plot points, explore alternative branches, and validate evolving narratives through an integrated story generation workflow. It further provides a structured interface to support efficient iteration over multiple story paths. Results from a user study with professional and semi-professional writers show that GraphStory reduces the effort of organizing narrative structures and better supports creativity and exploration compared to normal AI-based writing workflows.
\end{abstract}

\ccsdesc[500]{Computing methodologies~Computer vision}
\ccsdesc[500]{Human-centered computing~Information visualization}
\ccsdesc[500]{Computing methodologies~Natural language processing}
\ccsdesc[500]{Human-centered computing~Interactive systems and tools}

\keywords{AI-assisted creativity, Co-creative human–AI interaction, Graph-based storytelling, Interactive narrative editing, Event-based story modeling}

 
\begin{teaserfigure}
  \centering
  \includegraphics[width=\textwidth]{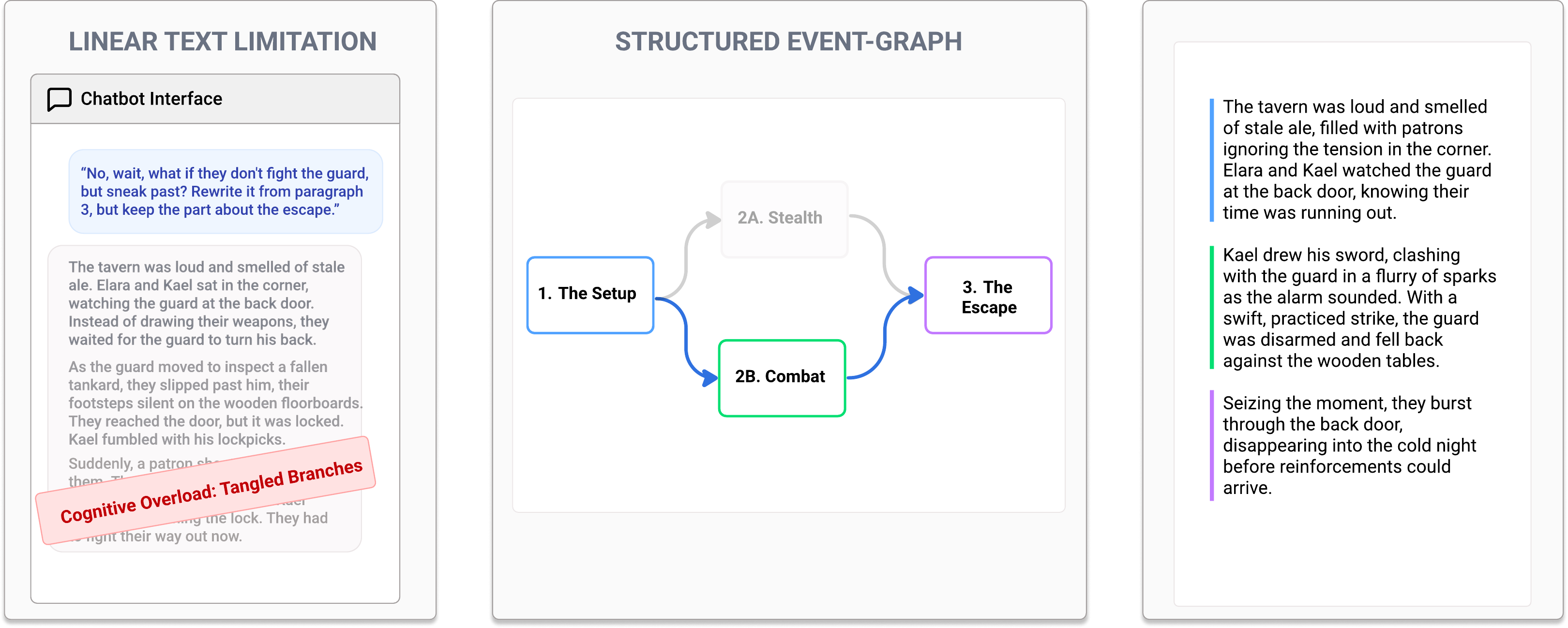}
  \caption{GraphStory concept and core workflow. (Left) Standard linear text interfaces often lead to cognitive overload when writers attempt to manage and rewrite branching ideas. (Middle) GraphStory introduces a structured event-graph to support explicit branching and multi-level editing, allowing users to efficiently structure their intent. (Right) The structured graph translates directly into a coherent narrative with clear visual provenance connecting the text back to the graph nodes.}
  \label{fig:teaser}
\end{teaserfigure}
\maketitle

\section{Introduction}

Story writing is a common activity across a wide range of domains, including literature, screenwriting, journalism, marketing, and interactive media such as games \cite{ryan2004narrative, dowd2015storytelling, zhang2024solving}. In these contexts, writers must construct engaging stories by organizing events, developing characters, and maintaining coherent narrative structures \cite{abbott2021cambridge, genette1980narrative}. Given its creative and cognitively demanding nature \cite{kellogg2008training, flower1981cognitive}, story writing has attracted significant attention from researchers seeking to design interactive applications to support this process \cite{kybartas2016survey, teleki2025survey, brown2020language}. In particular, recent advances in artificial intelligence, especially Large Language Models (LLMs) \cite{liu2024lost, huang2023advancing}, have enabled new systems that assist writers by generating text, providing suggestions \cite{lee2022coauthor, chung2022talebrush, yang2022re3, clark2018creative, ippolito2022creative}, and supporting different stages of the writing workflow \cite{yuan2022wordcraft, mirowski2023co, keskar2019ctrl, yao2019plan, rashkin2020plotmachines}.

Recent work has focused on enabling the generation of long-form, coherent narratives using a wide range of inputs \cite{chung2022talebrush, mirowski2023co, yang2022re3}. These systems can condition generation on prefixes or suffixes, as well as higher-level specifications such as topic, style, theme, and tone \cite{keskar2019ctrl, fan2018hierarchical, dathathri2019plug}. More advanced approaches further incorporate multimodal or structured inputs, including images, sketches, outlines, world settings, and character descriptions \cite{chung2022talebrush, yuan2022wordcraft, yao2019plan, rashkin2020plotmachines, huang2016visual}. By supporting diverse forms of input, these methods allow writers to better control the generated content and ensure alignment with their intent while maintaining fluency, coherence, and engagement \cite{teleki2025survey}. However, these approaches primarily support text production and control, and provide limited assistance for developing initial narrative flow. In this stage, writers must connect individual ideas into a coherent structure, reason about relationships between events, and iteratively explore alternative story developments. Existing systems often require writers to rely on sequential prompting or manual rewriting, making it difficult to externalize, organize, and refine evolving narratives. As a result, writers frequently struggle to connect disparate ideas and effectively iterate over different narrative directions.

Graph-based representations have been widely adopted in various domains to model complex and structured information, offering users an intuitive way to organize and explore relationships between elements \cite{herman2002graph}. Prior work has shown that graphs are effective for supporting sensemaking, planning, and knowledge organization, as they externalize connections and provide a clear overview of interconnected components \cite{novak2008theory, liu2008distributed, pirolli2005sensemaking}. Users often report that graph structures are more intuitive than linear formats when dealing with complex relationships, as they allow flexible navigation and direct controllability of individual elements \cite{andrews2010space}. These properties make graphs particularly well-suited for representing narrative structures, where events and plot points are inherently interconnected and often evolve in a non-linear manner. By explicitly capturing these relationships and supporting branching, graph representations provide a natural medium for developing and refining narrative flow.

In this paper, we present GraphStory, a writing support system that leverages a graph-based representation as an intermediate layer for both user interaction and LLM input. GraphStory enables writers to organize, connect, and iteratively develop their ideas into coherent narrative structures through an interactive graph interface (Fig.~\ref{fig:teaser}). The graph serves as structured context for the LLM, allowing the system to generate an instant output that are grounded in the evolving narrative and provide suggestion in improving connections between plot points. To support iterative development, GraphStory allows users to explore and compare alternative story branches, facilitating the validation and refinement of different narrative flows.

We evaluate GraphStory through a user study with professional and semi-professional writers. Our results show that GraphStory improves the writing experience by reducing the effort required to organize narrative structures and by better supporting creativity, exploration, and control compared to conventional LLM-based writing assistants.

In summary, our contributions are as follows:
\begin{itemize}
    \item We propose a graph-based interaction paradigm for story writing that uses a structured representation to support the development and organization of narrative flow while bridging user interaction and LLM reasoning.
    \item We design and implement GraphStory, an interactive system that enables writers to connect ideas, explore alternative story branches, and iteratively refine narratives with grounded LLM assistance.
    \item We present findings from a user study demonstrating that GraphStory reduces the effort of organizing narratives and better supports creativity, exploration, and control compared to conventional LLM-based writing workflows.
\end{itemize}

\section{Related Work}

\subsection{Narrative Writing with LLMs}

Recent advances in large language models (LLMs) have demonstrated strong capabilities in generating fluent, coherent, and contextually relevant long-form text \cite{chung2022talebrush, mirowski2023co, yang2022re3}. Both researchers and practitioners have recognized their potential for narrative writing tasks, including story generation, rewriting, and stylistic adaptation \cite{teleki2025survey, fan2019strategies}. Beyond casual use through chat-based interfaces, prior work has explored more structured and interactive systems that integrate LLMs into writing workflows, aiming to provide greater control and usability.

These systems aim to provide greater control over story generation by allowing users to specify a wide range of inputs that guide the output \cite{roemmele2015creative}. Early approaches rely on textual conditioning, such as prefixes or suffixes, where the model continues or completes a given piece of text. More recent work expands this by incorporating higher-level attributes, including topic, genre, style, theme, and tone, enabling writers to shape not only what is written but how it is expressed \cite{keskar2019ctrl, fan2018hierarchical, dathathri2019plug}. In addition, some systems introduce structured inputs such as outlines, world settings, and character descriptions, which help anchor the generation in a predefined narrative context \cite{yuan2022wordcraft, yao2019plan, rashkin2020plotmachines}. Beyond text, multimodal inputs, such as images, sketches, or videos, have also been explored to inspire or constrain storytelling \cite{huang2016visual}. By combining these diverse forms of input, these systems allow writers to exert finer-grained control over the generated content and produce narratives that better align with their intentions while maintaining coherence, fluency, and engagement.

Moreover, these systems seek to improve the user experience of story authoring by rethinking how writers provide input, moving beyond unrestricted natural-language prompting \cite{amershi2019guidelines}. Rather than relying on a single, free-form prompt, many interfaces introduce more structured and interactive mechanisms, such as editable forms, sliders, templates, and modular input fields \cite{montfort2006natural, ghaffari2025narrative, radwan2024sard, dhillon2024shaping}. These approaches allow writers to specify elements like characters, settings, and plot points incrementally, while making the effects of their choices easier to interpret and revise \cite{yuan2022wordcraft, rashkin2020plotmachines}. Some systems further support step-by-step workflows, guiding users through stages such as planning, drafting, and refinement, so that input can be provided in a more organized and iterative manner \cite{mirowski2023co, yao2019plan}. By making the input process more explicit and guided, these designs reduce ambiguity, lower the burden of prompt engineering, and provide writers with a clearer sense of control over the generated narrative.

Existing approaches to controllable text generation primarily focus on improving output quality \cite{teleki2025survey}, but provide limited support for developing and structuring narrative ideas. Writers often struggle to connect individual ideas into a coherent flow, explore alternative developments, and assess the impact of changes, as most systems rely on linear text and sequential prompting with little support for organizing evolving structures. To address this gap, our work shifts the focus to interaction design, introducing an interface that enables writers to structure, organize, and connect ideas before generation, and to receive outputs grounded in this structured representation.

\subsection{Graph-based Interaction}

Graph-based representations are widely used to model complex, structured information, providing an intuitive way for users to organize and explore relationships between interconnected elements. Prior research has demonstrated their effectiveness in tasks, such as sensemaking, planning, and knowledge management, where making relationships explicit helps users build and refine their understanding \cite{novak2008theory, liu2008distributed}. Compared to linear representations, graphs enable flexible navigation and direct manipulation, allowing users to focus on individual components while maintaining awareness of the overall structure \cite{andrews2010space, munzner2025visualization, hernando2018method}.

These characteristics make graphs particularly suitable for representing narrative structures. Stories are inherently composed of interconnected elements, including events, characters, and causal relationships, which often evolve in non-linear ways \cite{min2019modeling, padia2019system}. Existing work in narrative modeling has explored various graph-like structures, such as event networks and plot diagrams, to represent temporal sequences, causal dependencies, and character interactions \cite{yan2023narrative, li2018constructing, segel2010narrative, tanahashi2012design}. In domains like interactive storytelling and game design, branching structures are commonly used to model alternative story paths, enabling authors to design multiple possible developments and outcomes \cite{padia2019system, page1999hamlet}.

Graph representations enable a level of abstraction that is difficult to achieve with plain text, allowing writers to model high-level story components as nodes and refine them incrementally for both overview and detail-oriented editing. This facilitates restructuring narratives, exploring alternative configurations, and maintaining coherence across complex storylines through branching, reorganization, and iterative refinement. Building on this idea, our system extends prior work by treating the graph not merely as a visualization or planning aid, but as the central interaction medium for narrative writing. In GraphStory, the graph functions both as an interface for organizing and exploring story elements and as structured input to the LLM, tightly integrating user-defined structure with model-generated content to support more interactive and flexible development and validation of narrative flows.

\section{Formative Study}

To inform the design of our system, we conducted a formative study to better understand writers’ needs during narrative development. In particular, we focused on the stage where initial ideas are expanded into a coherent story, a process that involves organizing concepts, establishing connections between events, and iterating over alternative narrative directions. We also sought to explore how graph-based representations could support this process and what interaction mechanisms would be most intuitive and effective for writers. Based on this study, we aim to address the following research questions:

\begin{itemize}
    \item \textbf{RQ1:} What challenges do writers encounter when developing initial ideas into a coherent narrative structure?
    \item \textbf{RQ2:} How do writers expect to interact with graph-based representations to support narrative development and exploration?
\end{itemize}

\subsection{Participants}

We recruited five participants (aged 18–22; 2 male, 3 female), all of whom were students majoring in writing-related fields, including screenwriting, literature, and copywriting. We refer to them as FP1–FP5 throughout the paper. All participants were experienced in writing and had prior experience using LLMs to support their work. Participants provided informed consent and received \$5 as compensation for their participation.

\subsection{Study Design and Procedure}

We conducted a two-part formative study to investigate writers’ practices and challenges during narrative development. First, we carried out a semi-structured interview to understand how participants transform initial ideas into complete stories, focusing on their workflows, strategies for organizing ideas, and difficulties in connecting events and iterating over narrative directions.

In the second part, we developed a simple prototype that represents a story as a graph, where nodes correspond to narrative elements and edges represent progression and branching. Participants interacted with this interface to modify and reorganize a familiar story, and they could submit their edits to a large language model (LLM) to generate updated versions reflecting those changes. They followed a think-aloud protocol \cite{ericsson2017protocol} while performing these tasks, allowing us to observe how they engaged with the graph and used LLM feedback.

Each session was conducted in a private room. Participants were first briefed on the study goals, then completed the initial interview, followed by a 30-minute interaction with the prototype with researcher support. Finally, they participated in a follow-up interview to provide feedback on the system’s usability and usefulness.

\subsection{Analysis}

We conducted a thematic analysis \cite{braun2021thematic} of the collected data, including the initial interviews, interaction sessions, and follow-up interviews. We adopted an inductive approach to identify recurring patterns related to participants’ writing practices, challenges, and interactions with the graph-based prototype. The analysis process involved iterative coding and refinement, with multiple rounds of review and discussion among the authors to consolidate themes. Subjectivity was mitigated through this iterative process and multi-stage review.

\subsection{Narrative Flow Development Challenge}

\textbf{Iterative Process.}
Participants consistently described narrative development as a highly iterative process. Rather than following a fixed sequence, they often explored multiple possible directions, revised earlier ideas, and refined story elements over time, involving frequent backtracking and comparison between alternatives when evaluating how different plot developments might influence the overall story. However, participants noted that existing tools provide limited support for such iteration, often requiring them to manually track or rewrite different versions. For example, FP1 shared that she “often struggles to trace back old ideas when using tools like ChatGPT,” making it difficult to revisit prior thoughts, while FP3, a screenwriting student, explained that he “usually prepares multiple story flows to compare and present to teacher,” highlighting the need to manage parallel narrative paths. This makes it difficult to efficiently explore and manage multiple narrative trajectories.

\textbf{Coherent Connections Between Narrative Elements.}
Maintaining coherence across narrative elements emerged as a key challenge. Participants reported difficulty in establishing clear relationships between events, particularly when integrating new ideas into an existing structure, often needing to reason about causal, temporal, and thematic connections. For instance, FP2 explained that they “often think of some key events first and then try to connect them,” highlighting the effort required to link story components meaningfully, while FP5 noted that “initial ideas are often not coherent,” requiring further refinement to achieve consistency. As a result, ensuring that the story flows logically and cohesively across different parts demanded significant cognitive effort.

\textbf{Content Length Guarantee.}
Participants highlighted the need to ensure that a story contains an appropriate amount of content depending on its intended form, such as a short story, script, or longer narrative. This often requires adding, expanding, or condensing events to meet expectations for pacing and completeness, as writers adjust the number of narrative elements to ensure the story feels sufficiently developed without being overly sparse or unnecessarily detailed. However, managing this balance can be challenging, as it requires maintaining awareness of both the overall structure and the contribution of individual events. For example, FP5 noted that they “may need to add more detail to ensure the content scope even when the current plot feels complete,” which can lead to introducing new events without disrupting the existing structure or, in some cases, requiring changes to the overall flow.

\subsection{Graph-based Interaction}

\textbf{Comprehensive Overview with Multi-level Detail.}
Participants reported that the graph representation provides a more comprehensive view of the narrative compared to linear text. By organizing story elements as nodes and connections, the graph allows writers to simultaneously perceive the overall structure while inspecting individual components, supporting both high-level understanding and localized editing. For example, FP2 explained that “the graph view of the full story helps me know where to add more events and trace what it will affect,” highlighting how the representation supports reasoning about structural impact. This dual perspective enables participants to move more effectively between abstract story flow and detailed narrative elements than with paragraph-based formats.

\textbf{Flexible Exploration of Narrative Flow.}
Participants found it easy to explore and test different narrative flows using the graph structure. In particular, modifying edges to change progression or branching allowed them to quickly reorganize the story and evaluate alternative directions. This interaction was perceived as more efficient and intuitive than rewriting text, as it enabled users to focus on structural changes without altering the entire narrative. For example, FP1 highlighted that adjusting connections between nodes made it easier to try out different story paths, while FP4 noted that “adjusting the graph and seeing the output from AI inspires how to write his own story,” emphasizing how rapid feedback supports creative exploration. As a result, the graph representation supports rapid experimentation and helps writers systematically explore and compare different story development.

\section{GraphStory}
\begin{figure*}[t]
  \centering
  \includegraphics[width=\textwidth]{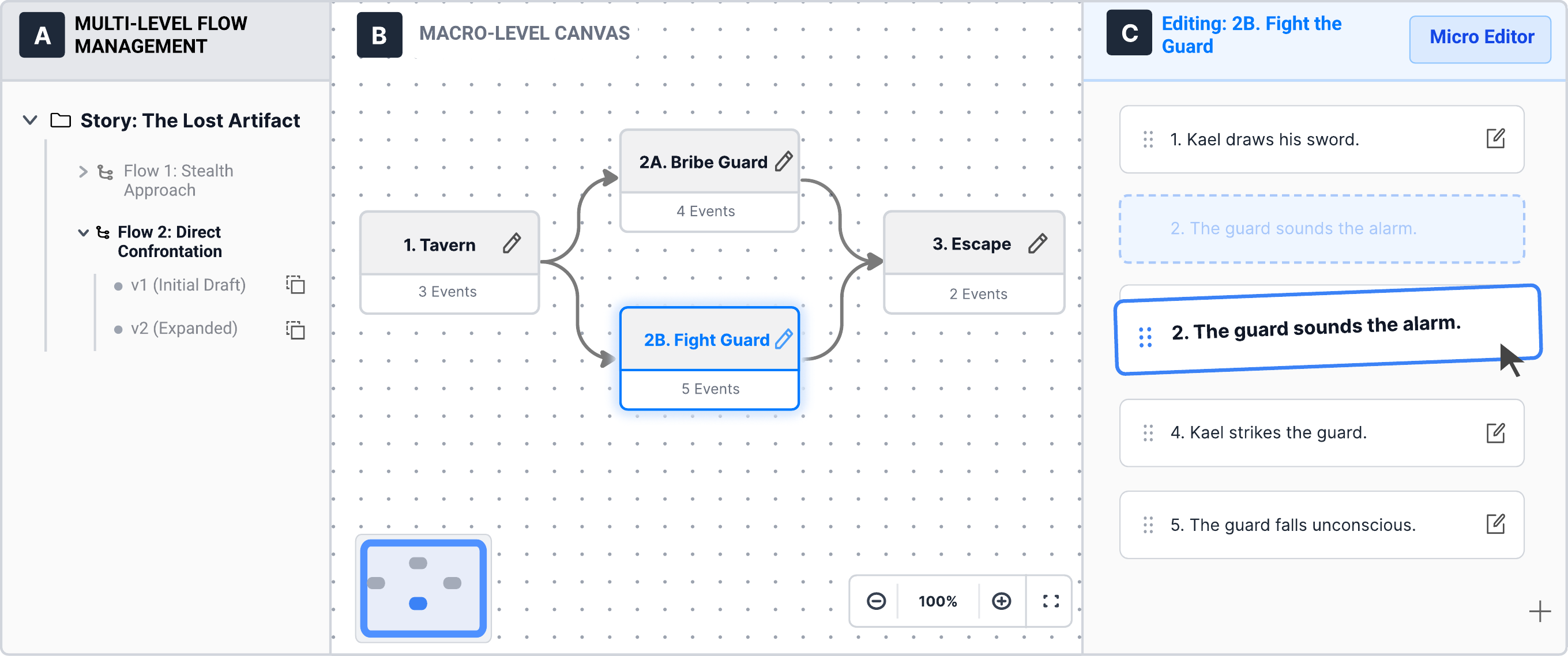}
  \caption{The main interface of GraphStory. (A) The Multi-level Flow Management panel organizes variations by Story, Flow, and Version. (B) The Macro-level Canvas provides a comprehensive overview of the narrative structure, allowing users to build and branch storylines using interconnected nodes (chunks). (C) The Micro-Editor enables fine-grained, drag-and-drop manipulation of individual events within a selected chunk.}
  \label{fig:interface}
\end{figure*}
\subsection{System Overview}
Based on insights from our formative study, we propose GraphStory, a grounded system designed to support writers in developing and refining their ideas through structured yet flexible exploration. The system comprises three core components: an Event-Graph Constructor, a Story Generator, and a Multi-level Flow Management module. The Event-Graph Constructor transforms user inputs into a sequence of nodes, providing a comprehensive representation that facilitates the organization and progressive development of narrative ideas. Building on this structure, the Story Generator functions as a human-in-the-loop module that produces provisional narrative outputs, offering writers reference points to validate emerging storylines, compare alternative flows, improve coherence, and manage content scope. Complementing these components, the Multi-level Flow Management module enables iteration across different levels of organization, allowing users to explore, compare, and refine narrative flows to identify those that best align with their intent.

\subsection{Event-Graph Constructor}

\begin{figure*}[t!]
  \centering
  \includegraphics[width=\textwidth]{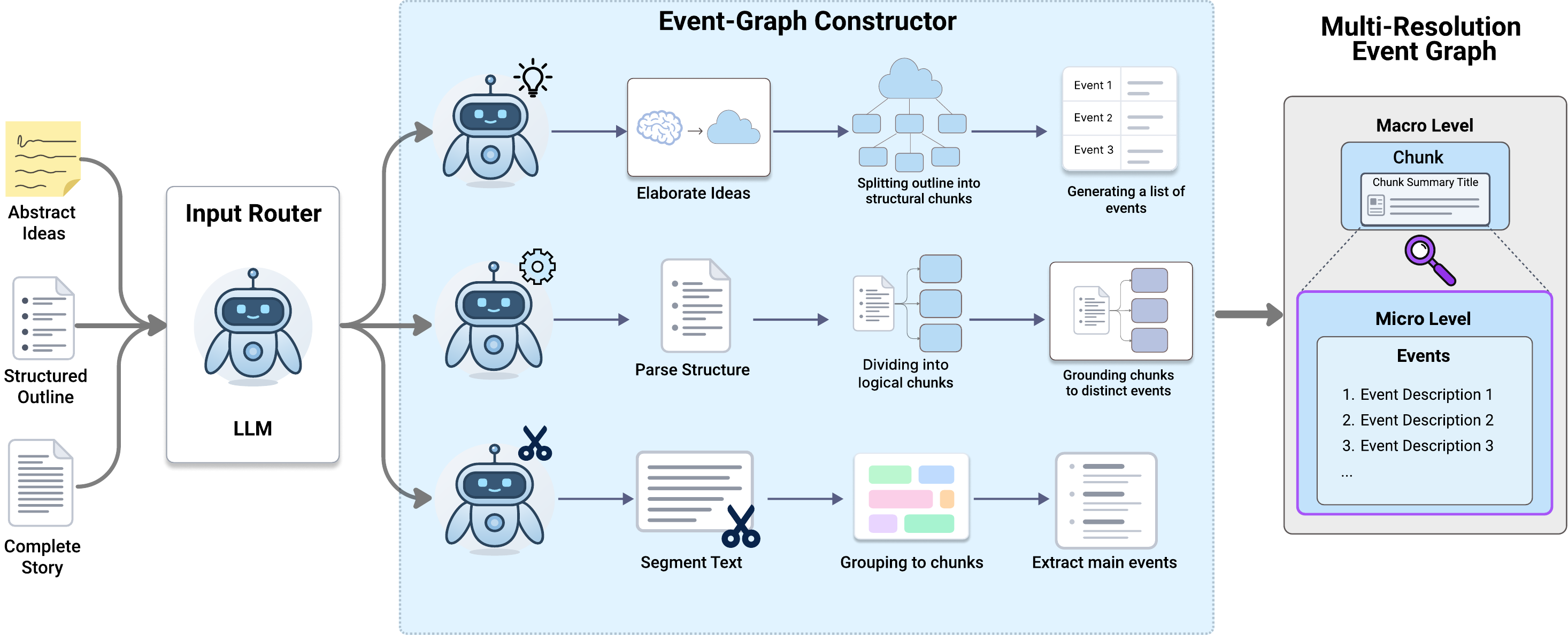}
  \caption{The Event-Graph Constructor pipeline. The Input Router processes three distinct forms of input (Abstract Ideas, Structured Outline, and Complete Story) via specialized workflows. Each pathway translates the input into structural chunks at the Macro Level and distinct narrative events at the Micro Level to build the event-graph.}
  \label{fig:event_constructor}
\end{figure*}

The Event-Graph Constructor is a core module designed to transform user-provided content into a structured event graph that represents the narrative flow (Fig.~\ref{fig:event_constructor}). It receives input in three forms: abstract ideas, structured outlines, and complete stories, each of which can be provided via text prompts or uploaded documents (e.g., PDF or Word files). Users explicitly choose the type of input, allowing the module to adapt its processing strategy to best fit the provided material. The output is a sequence of nodes that captures the key narrative events and organizes them in a hierarchical format, enabling writers to visualize and refine the story structure effectively.

To process abstract ideas, the system constructs a fine-grained outline from user input through a two-stage procedure. It first generates a preliminary outline that organizes the ideas into high-level sections, referred to as chunks, which represent major narrative units. It then elaborates each chunk by generating a corresponding list of events that capture key plot points and instantiate the narrative content, transforming initial concepts into a structured sequence that supports early-stage exploration and development.

For structured outlines, the system follows a similar process by converting the input into a more detailed representation. It divides the outline into high-level sections, refered to as chunks, based on its inherent structure and then generates a corresponding list of events for each chunk grounded in its content. This expands the outline into a coherent sequence of key narrative points while preserving the original organizational intent. 

For complete stories, the system first segments the text into paragraphs and determines an appropriate grouping size to combine consecutive paragraphs into chunks, forming higher-level narrative units that maintain coherence. Each chunk is then parsed to identify its main events, resulting in a structured and detailed representation of the story’s narrative flow.

Across all input types, the resulting sequence of nodes serves as the initialization of the event graph on the current flow. The interface provides a multi-resolution editing paradigm to manage both overarching plot and detailed narrative elements (Fig.~\ref{fig:interface}). At the macro-level, the graph offers an overview of the structure (Fig.~\ref{fig:interface}B): nodes (representing chunks) display a concise summary title and are connected by structural arrows depicting story progression. Writers can organize the plot by connecting nodes or extending paths to illustrate branching storylines during brainstorming. At the micro-level, writers can utilize semantic zooming to expand a node and inspect its numbered list of events. By interacting with a specific node, users open a dedicated event editor (Fig.~\ref{fig:interface}C). Here, they can utilize direct manipulation to drag-and-drop, reorder, edit, delete, or manually add individual events strictly within the boundary of that specific chunk. Together, these interactions enable flexible, iterative manipulation of the narrative, supporting both high-level restructuring and fine-grained editing.

\subsection{Story Generator}
\begin{figure*}[t]
  \centering
  \includegraphics[width=\textwidth]{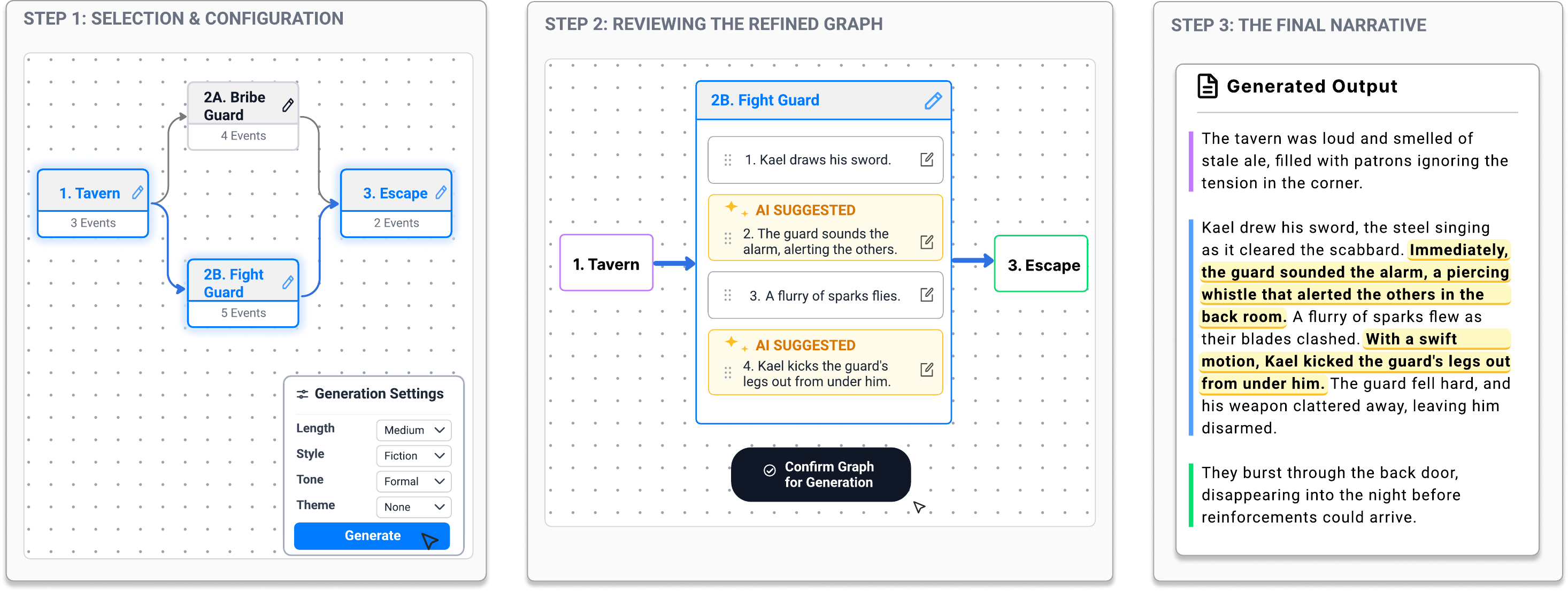}
  \caption{Interactive story generation workflow. (Step 1) Users select an origin node and path to configure the generation queue. (Step 2) The system operates a 2-stage generation, resulting in a refined graph where original events are preserved and AI-suggested intra-chunk events are highlighted in yellow for human-in-the-loop review. (Step 3) After user confirmation, the system generates the final coherent narrative, with visual provenance matching the text to the AI-suggested events. (Note: The yellow text highlighting in the Generated Output panel is for visualization purposes in this paper only, demonstrating how specific AI-suggested events translate into the final prose.)}
  \label{fig:workflow}
\end{figure*}

The story generator module is designed to produce a refined narrative from user-selected chunks, supporting validation of a chosen storyline while offering guidance on improving coherence and managing content scope. To initiate generation for a specific narrative branch, users enter a selection mode by right-clicking an origin node to anchor the path, followed by sequentially left-clicking subsequent nodes to define the exact narrative chain (Fig.~\ref{fig:workflow}, Step 1). As nodes are selected, they illuminate and are connected by blue arrows, visually distinguishing the active generation queue from the underlying structural graph. If a user wishes to alter the sequence or cancel the selection, they simply click anywhere on the empty graph space to reset the path. In addition, users specify a target content length and may optionally configure attributes such as writing style, tone, or theme, which guide subsequent generation without constraining manual edits.

The generation procedure constructs a new sequence of events in two stages with a human-in-the-loop confirmation. In the first stage, the system operates at the intra-chunk level by adding new events to each selected chunk to improve local coherence and, when necessary, better align with the desired content scope, without modifying existing events so as to preserve the user’s original intent. In the second stage, the system considers the sequence of chunks at a global level, modifies added events, and inserts new events to improve transitions and overall flow across chunks. These additions are designed to provide supportive suggestions rather than enforce correctness, leaving users in control of refining or modifying the narrative.

Following this two-stage generation, the system presents the newly mixed sequence of original and AI-added events as a refined graph, utilizing the exact same visual interface as the baseline flow (Fig.~\ref{fig:workflow}, Step 2). To ensure clear visual provenance and maintain user agency, all AI-generated events are highlighted in a distinct color, instantly differentiating them from the writer's original ideas. This visual and interactive consistency drastically reduces the user's cognitive load; users see the AI's inter-chunk and intra-chunk suggestions integrated directly into the familiar node-and-event-list structure. Importantly, interaction is not restricted to the AI-generated content. Using the standard event editor, users can easily examine, modify, or reject the color-coded suggestions, while also retaining the ability to seamlessly edit any original events across all chunks. This comprehensive control allows authors to iteratively adjust the entire structure through direct manipulation, smoothing the narrative flow exactly to their preferences. Once users are satisfied with this refined graph and provide confirmation, the system inputs the finalized structure into the underlying LLM (specifically, GPT-4o \cite{hurst2024gpt}), incorporating user-specified configurations such as style and theme to generate a complete narrative text (Fig.~\ref{fig:workflow}, Step 3). To support user evaluation and explainability, the interface features a mapping mechanism that links chunks to their corresponding generated text segments. The resulting output provides a smooth and natural reading experience, enabling writers to intuitively sense the story flow while easily tracing how the LLM translated their events into prose.

\subsection{Multi-level Flow Management}

The Multi-level Flow Management module (Fig.~\ref{fig:interface}A) provides structured organization and version control for story development, supporting multiple levels of narrative exploration. It is designed around three levels: story, flow, and version. When a user begins their work by providing input to the Event-Graph Constructor, a new story-level unit is created. This unit serves as the overarching container, encompassing all variations of event graphs and story iterations that the writer generates.

Within a story unit, each flow represents a distinct progression of the narrative. Flows are particularly useful when writers wish to explore entirely different story directions or when an existing graph becomes too complex due to an excessive number of chunks and events. Each flow captures the structure and sequence of a particular variation, allowing writers to experiment without losing prior work.

A version is created whenever a story is generated from a selected set of chunks, capturing the refined sequence of events produced during the story generation phase along with the resulting narrative text. Rather than storing the original chunks alone, each version preserves this enhanced event structure as a concrete representation of the developed storyline. Writers can reuse a version by initializing a new flow with its associated graph, using the refined sequence of events as the starting point for further exploration and modification. This organization, spanning story, flow, and version, provides flexible control over the creative process, supporting iterative development, exploration of alternatives, and efficient reuse of previously generated structures.

\section{User Study}

We conducted a study to evaluate our system in terms of functionality, usability, and its impact on reducing cognitive load for writers. The study aimed to understand how effectively the system supports writers in managing complex narratives, generating coherent stories, and iteratively refining their ideas. By examining these aspects, we sought to assess the overall effectiveness of the system in facilitating both creative and cognitive aspects of story writing.

\subsection{Participants}
We recruited 16 students aged 18 to 22 (8 male, 8 female) to participate in the study. We refer to them as UP1–UP16 throughout the paper. All participants were majoring in fields related to writing and possessed advanced writing skills. None of the participants had taken part in any prior study involving our system. All participants reported being familiar with AI-based tools for writing. Informed consent was obtained from each participant, and they were compensated at a rate of \$5 per hour for their participation.

\subsection{Study Design and Procedure}

\begin{figure*}[t]
  \centering
  \includegraphics[width=\textwidth]{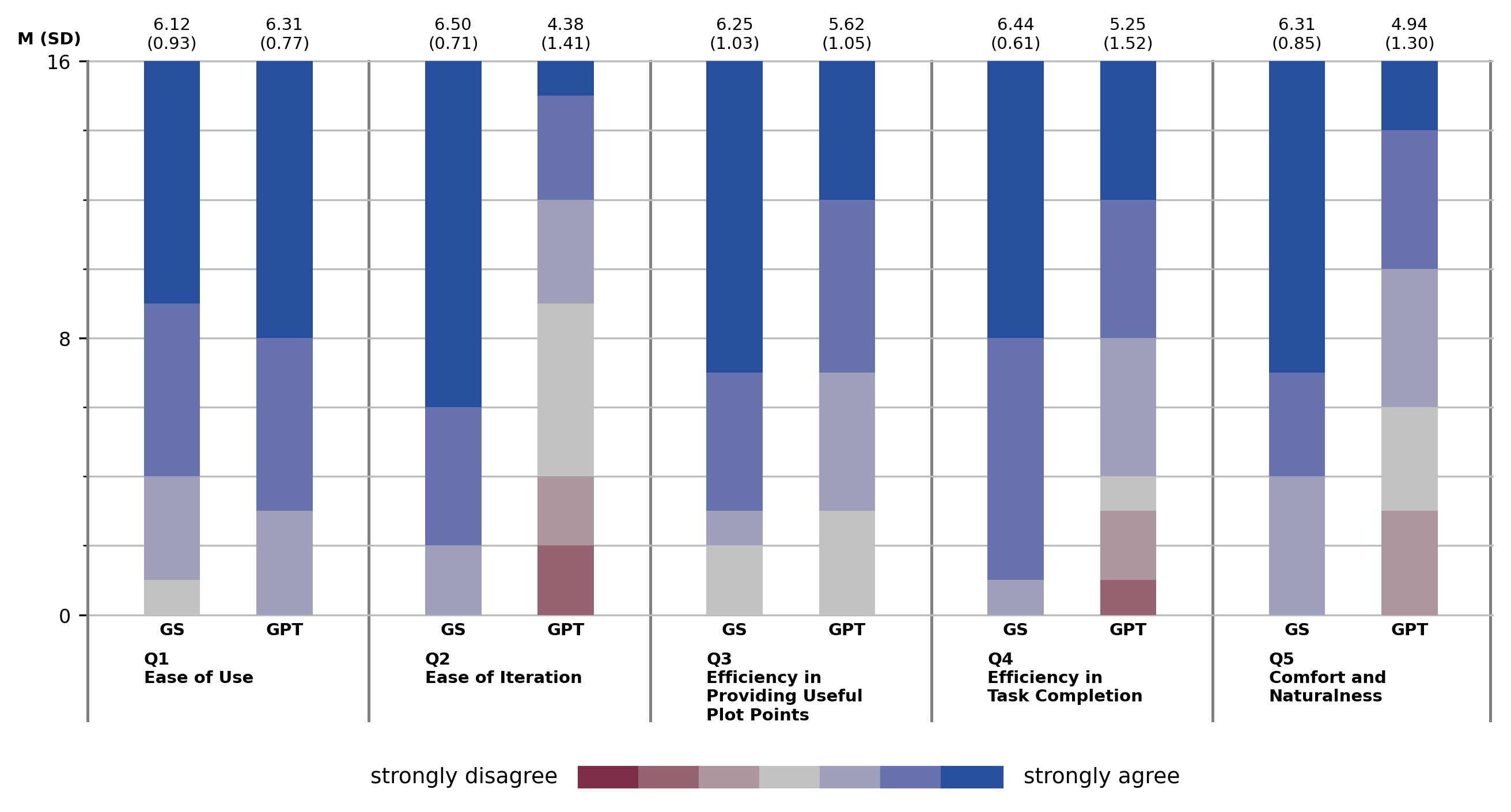}
  \caption{Questionnaire results on the 7-point Likert scale from "Strongly Disagree" to "Strongly Agree" in User Study to compare between GraphStory (GS) and ChatGPT (GPT). The $x$-axis represents different questions, covering various aspects such as system usability and interaction experience. The $y$-axis shows the Likert scale ratings provided by participants for two systems.}
  \label{plot:comp}
\end{figure*}

Each participant took part in the study individually in a private room, with one of the authors present to provide guidance and support as needed. Each session lasted approximately one hour. Participants were asked to complete a single story-writing task using a set of provided ideas.

Participants first wrote a story using ChatGPT and then completed the same task using our system. They were instructed to follow a think-out-loud protocol throughout the process, verbalizing their thoughts, strategies, and reasoning as they interacted with the tools. The entire session was recorded to capture both screen activity and verbalizations, allowing detailed analysis of participants’ interactions and cognitive processes during story creation.

During the session, additional feedback was collected as participants worked on the tasks, providing immediate insights into their experience. After completing both tasks, participants filled out evaluation forms using a Likert-scale format based on predefined metrics. In addition, they completed the NASA Task Load Index (NASA-TLX) questionnaire \cite{hart1988development} to assess perceived workload across multiple dimensions. Finally, semi-structured interviews were conducted to gather more detailed qualitative feedback about their experience, preferences, and perceptions of the system compared to ChatGPT. This procedure enabled a comprehensive assessment of usability, functionality, and cognitive support in a controlled and consistent setting.

\subsection{Quantitative Results}

For the statistical analysis, we used Wilcoxon signed-rank tests~\cite{woolson2007wilcoxon} to analyze the data. Statistical significance was established at $p < 0.05$ for all analyzes.

\subsubsection{Interaction experience and generated content evaluation}
A questionnaire using a 7-point Likert scale \cite{joshi2015likert} (as shown Fig. \ref{plot:comp}) is used to assess the usability of our system compared to ChatGPT for this task. While chat-based interaction provides slightly greater ease of use than our graph interface (Q1, $p=0.26$), our system better fulfills its purpose as a tool for writers to organize, brainstorm, and iteratively develop narrative ideas. It outperforms ChatGPT in terms of ease of iteration (Q2, $p < 0.001$), task efficiency (Q4, $p < 0.01$), and user comfort (Q5, $p < 0.001$). The slight increase in user ratings (Q3, $p < 0.0 5$) for the plot points generated by both systems suggests that the story generation module performs better than relying on ChatGPT alone.

\begin{figure}[t]
  \centering
  \includegraphics[width=\columnwidth]{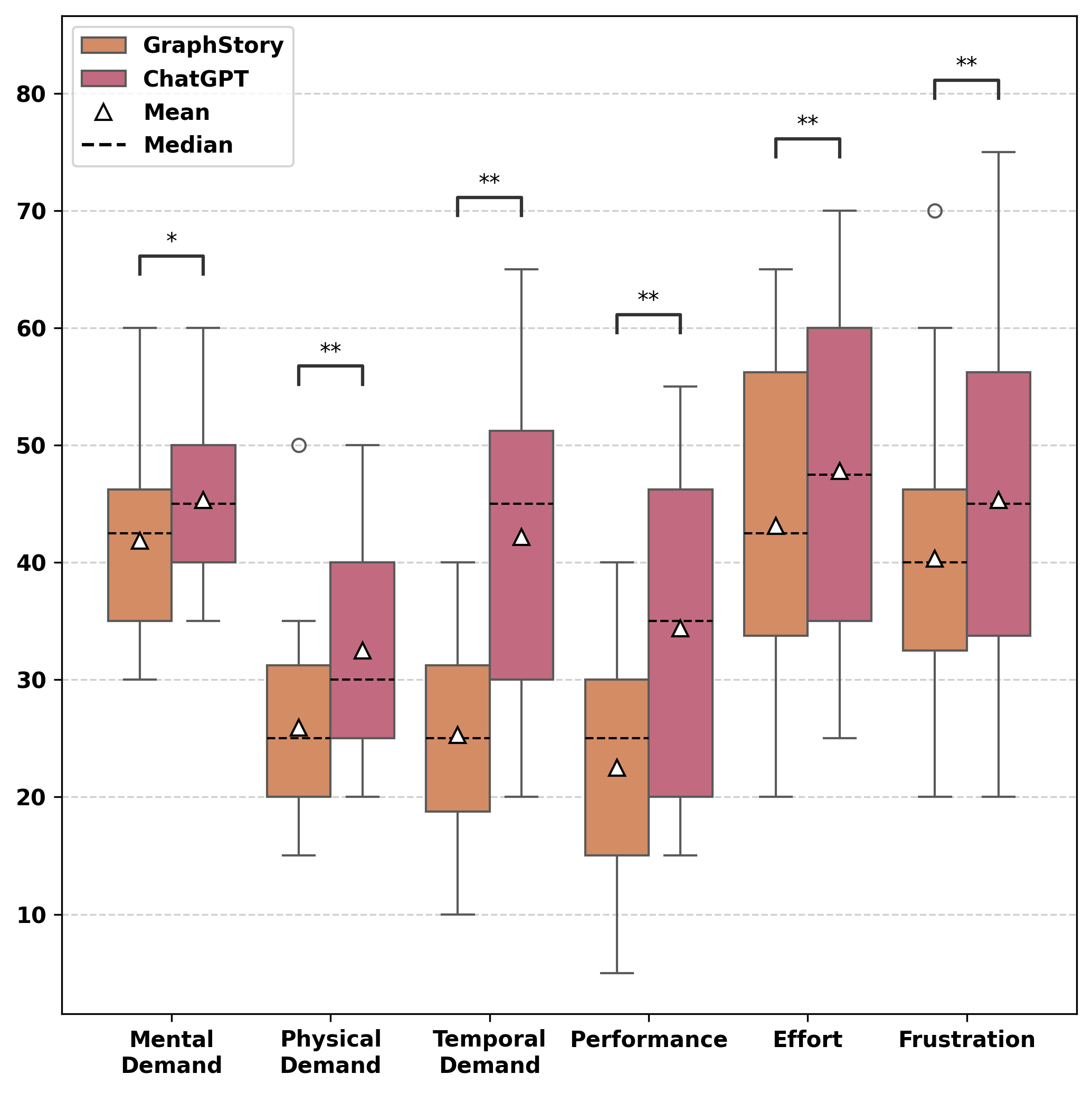}
  \caption{Task load results in NASA-TLX from User Study (the lower the better). The $y$-axis displays the six NASA-TLX subscales. The $x$-axis shows the corresponding scores. (* indicates $p < .05$, ** indicates $p < .01$)}
  \label{plot:tlx}
\end{figure}

\subsubsection{Task load}
NASA-TLX results (Fig.~\ref{plot:tlx}) indicate that GraphStory reduces perceived workload across multiple dimensions. Participants reported lower mental, physical, and temporal demands, while also rating their performance more favorably compared to the baseline condition. In addition, frustration levels were consistently lower, suggesting a more comfortable and manageable interaction experience.

These findings imply that the graph-based representation, together with the multi-level flow management mechanism, helps users better organize and navigate narrative elements. By providing a clearer structural overview and supporting incremental exploration, the system reduces cognitive effort and time pressure during story development. As a result, users can focus more on creative decision-making rather than on managing complexity, leading to a more efficient and less stressful writing process.

\begin{figure}[t]
  \centering
  \includegraphics[width=\columnwidth]{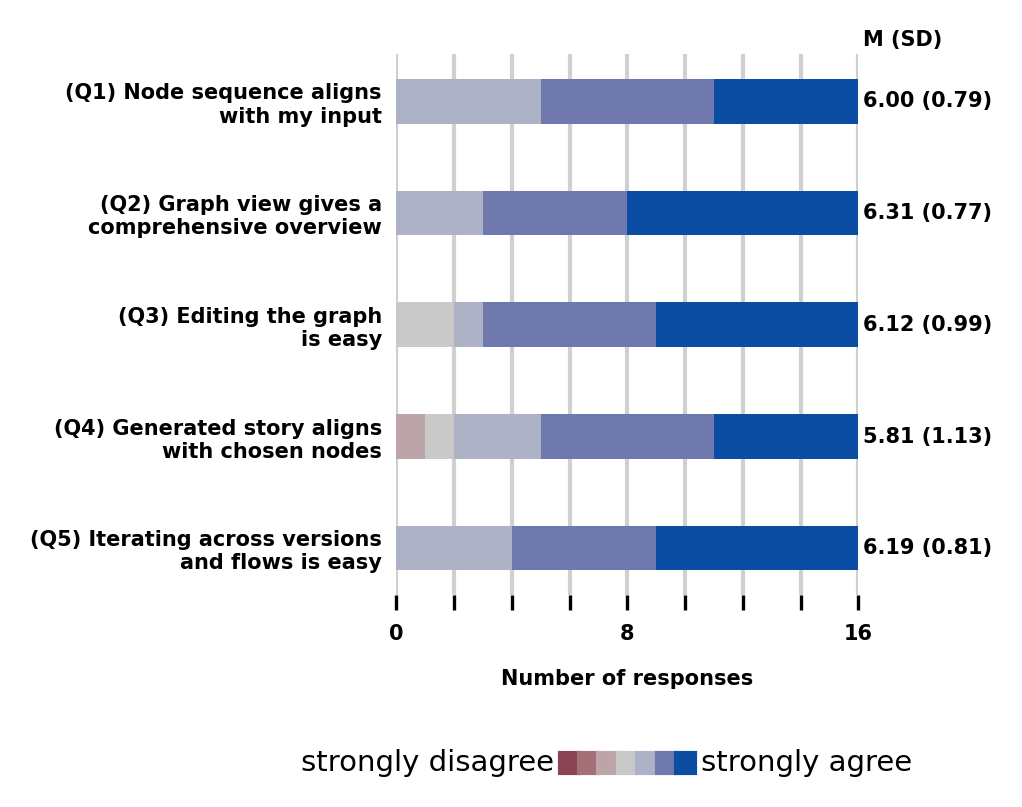}
  \caption{Questionnaire results on the 7-point Likert Scale from User Study to evaluate functionality proposed in GraphStory. Ratings are collected on a scale from “strongly disagree” to “strongly agree”.}
  \label{plot:eval}
\end{figure}

\subsubsection{Functionality evaluation}
We used a 7-point Likert-scale \cite{joshi2015likert} questionnaire (Fig. \ref{plot:eval}) to evaluate key functionalities of GraphStory, including node sequence generation (Q1), the graph representation and interaction engine (Q2–Q3), the generated story (Q4), and the multi-level management system (Q5). The results show consistently high ratings across most questions, indicating that participants found the system effective and easy to use. Overall, these findings suggest that GraphStory provides strong support for structured and flexible narrative development.

\subsection{Qualitative Results}

\textbf{Comprehensive Graph Representation.} Participants consistently emphasized that the event graph offered a clear and well-structured view of their narratives. By organizing story elements from high-level concepts down to detailed events, the representation made relationships and progression more transparent across multiple levels of abstraction. This hierarchical structure enabled writers to maintain an overview of complex storylines while also identifying gaps or inconsistencies. Several participants highlighted the ease of navigation, noting that they could quickly locate and edit specific parts of the story without rereading large portions of text. Others pointed out that the graph supported early-stage ideation by allowing them to capture and organize emerging ideas without committing to fully written prose.

\textbf{Acceleration of the Iterative Process.} The system was widely perceived as speeding up the iterative writing process. Participants described how immediate outputs, such as updated graphs or draft text, allowed them to rapidly test narrative variations and refine story structures. This reduced the effort typically required to track and revise interconnected story elements, enabling greater focus on creative decisions. The ability to explore multiple narrative paths before committing to one direction was particularly valued. In addition, compared to traditional text-based tools, the system’s structured approach to managing revisions made it easier for participants to revisit and modify earlier ideas without losing track of changes.

\textbf{Generation of Story Ideas.} The story generation module played a key role in supporting creativity by introducing new plot points and connections. Participants found that the generated content often surfaced possibilities they had not initially considered, helping them expand and enrich their narratives. By illustrating how different story components could be linked, the system encouraged exploration of alternative directions while preserving a sense of authorial control. Some participants noted that longer generated outputs were especially useful for uncovering additional narrative opportunities, as they provided more detailed and varied suggestions.

\textbf{Challenges with AI-Generated Content.} Despite these benefits, participants also reported difficulties when interacting with AI-generated suggestions. In particular, an overabundance of newly introduced events sometimes led to confusion and disrupted the intended narrative flow. Deciding which suggestions to incorporate could become challenging, especially when outputs diverged significantly from the original storyline. Participants also observed that different generated versions, while sharing similar high-level plots, could vary substantially in detail, making it harder to maintain coherence across iterations. These issues suggest that while generative support enhances creativity, it can also increase cognitive load when not carefully balanced.

\section{Discussion}

\textbf{Structured Graph Representations Bridge User Intent and AI Generation.} A core aspect of GraphStory's design is the use of a structured event graph as a transparent, editable intermediate layer between user intent and LLM generation. Rather than treating prompts as opaque, one-shot inputs as in chat-based interfaces~\cite{mirowski2023co}, this representation aligns with human-centered AI principles~\cite{amershi2019guidelines} by making the mapping from intent to output inspectable and revisable at multiple levels of granularity. Supporting editing at both the macro (chunk) and micro (event) levels mirrors the divergent-convergent creative process~\cite{frich2019mapping, chung2022talebrush, shneiderman2007creativity}; systems that constrain writers to a single level of abstraction disrupt this natural oscillation, whereas GraphStory's two-layer model explicitly accommodates both open-ended structuring and fine-grained refinement. Our quantitative results support this: GraphStory clearly outperformed ChatGPT in ease of iteration, task efficiency, and user comfort. Furthermore, functionality evaluations showed that participants highly rated the graph representation for providing a comprehensive overview and for its ease of editing. GraphStory's three-level organization (story, flow, version) further addresses the challenge of tracking diverging narrative directions simultaneously-something linear chat interfaces fundamentally cannot support~\cite{sterman2022towards}. Participants confirmed this in qualitative feedback, noting that the system's structured approach to managing revisions made it easier to revisit and modify earlier ideas without losing track of changes, and that iterating across versions and flows was straightforward.

\textbf{Visual Provenance Builds Trust in AI-Generated Content.} Highlighting AI-suggested events in a distinct color and mapping them directly to generated prose gave participants greater confidence and editorial control. In qualitative feedback, participants consistently emphasized that the visual distinction between original and AI-generated content allowed them to quickly assess, accept, or reject suggestions without losing track of their own contributions. These observations are consistent with findings in explainable AI research~\cite{amershi2019guidelines, miller2019explanation, liao2020questioning}, which show that provenance mechanisms rendering AI contributions legible are essential for sustaining authorial agency in co-creative workflows~\cite{lee2022coauthor}. Our functionality evaluation further supports this, as participants favorably rated the alignment between generated stories and their selected nodes, suggesting that the two-stage generation process with visual provenance provides a meaningful degree of controllability over AI outputs.

\textbf{Balancing AI Agency and Writer Control.} A recurring tension in GraphStory is calibrating the boundary between AI contribution and writer autonomy~\cite{buccinca2021trust}. When the story generator proposes inter-chunk events, it inevitably introduces narrative directions the writer did not intend. Our qualitative findings confirm this duality: participants valued creative suggestions and reported that generated content surfaced possibilities they had not initially considered, but they also noted that an overabundance of AI-added events was sometimes disruptive and could increase cognitive load. This tension is consistent with broader concerns about over-reliance on generative systems~\cite{mirowski2023co, yuan2022wordcraft, buccinca2021trust} and reflects a fundamental co-creative design challenge-systems contributing too little provide insufficient scaffolding, while those contributing too much erode authorial identity~\cite{lee2022coauthor}. Notably, while survey responses indicated that chat-based interaction was perceived as slightly easier to use initially, this did not translate into better creative outcomes. This suggests that the modest learning curve of graph-based interaction is successfully offset by its structural advantages. Future systems should explore adaptive mechanisms that modulate AI contribution based on writer preferences and the current creative phase.

\textbf{Broader Applicability and Ethical Considerations.} While GraphStory targets short-to-medium narrative writing, the underlying paradigm of using a structured graph as an editable intermediate layer extends to screenwriting~\cite{mirowski2023co}, game dialogue trees~\cite{ryan2004narrative}, and interactive fiction~\cite{kybartas2016survey}. Recent LLM-powered authoring tools confirm the broader value of structured representations for bridging user intent and system execution~\cite{gunturu2025mapstory, huang2025sketchgpt}. Ethical considerations also arise when applying this paradigm more broadly: authorship attribution becomes contested when AI substantially shapes narrative content~\cite{mirowski2023co}, and models trained on large fiction corpora may encode cultural biases in story archetypes or character demographics~\cite{teleki2025survey} that writers absorb without awareness.

\section{Limitations \& Future Work}

Despite GraphStory's demonstrated benefits, several limitations point to concrete future directions.

\textbf{Non-determinism in generation.} AI-generated content is stochastic: the same input can produce substantially different event sequences across runs. While variability can introduce novel ideas~\cite{chung2022talebrush, yuan2022wordcraft}, it makes it difficult for writers to isolate the precise effects of small edits, disrupting iterative workflows~\cite{mirowski2023co}. Future work should explore temperature-aware generation controls that let writers explicitly trade variability for consistency depending on their creative phase.

\textbf{Graph-narrative alignment.} Generated prose does not always faithfully reflect the event graph-text may conflate chunks, omit events, or introduce content outside the specified structure. This is partly inherent to current LLMs~\cite{yang2022re3, teleki2025survey}, which struggle to maintain structural fidelity over long outputs. While GraphStory's two-stage generation mitigates this, future approaches should explore constrained decoding strategies that ground narrative outputs more reliably in the specified event sequence.

\textbf{Participant population.} Our study recruited 16 students aged 18–22 in writing-related fields, limiting generalizability to professional writers such as novelists or screenwriters, who may have different workflows and evaluation criteria~\cite{mirowski2023co}. Our current metrics may also not fully capture priorities such as stylistic voice or narrative coherence. Future work should conduct longitudinal studies with professionals in authentic production contexts.

\textbf{Baseline and evaluation scope.} Using ChatGPT as the sole baseline, while representative of common practice, does not isolate the contribution of graph-based structuring against dedicated outline-based tools~\cite{yao2019plan, rashkin2020plotmachines} or structured co-authoring systems~\cite{lee2022coauthor, yuan2022wordcraft}. Furthermore, GraphStory was tested on moderately complex stories; at scale, with dozens of nodes or many parallel flows, the canvas may become visually cluttered and the cognitive benefit of graph overview may diminish~\cite{andrews2010space}. Future work should include ablation comparisons against graph-free baselines and investigate adaptive layout algorithms with progressive disclosure mechanisms to support large-scale narrative graphs.

\section{Conclusion}

We presented GraphStory, an interactive writing support system designed to facilitate creative story writing through a graph-based representation of narrative elements. By providing a clear overview of story structure and supporting the organization and connection of events, GraphStory enables writers to explore alternative narrative paths, validate evolving storylines, and iterate efficiently over multiple story flows. Evaluation results indicate that the system reduces the cognitive effort required to manage complex narratives, accelerates the iterative writing process, and supports creativity by offering new story ideas, while maintaining user control over narrative development.

These findings highlight the benefits of integrating visual, structured representations with AI-assisted generation in creative writing workflows. While limitations remain, such as variability in AI-generated content and occasional misalignment with longer narratives, GraphStory demonstrates the potential of graph-based tools to enhance both the creative and cognitive aspects of storytelling. In future work, we aim to improve the consistency of generated content and further strengthen the connection between the graph and story output, continuing to support writers in exploring, refining, and realizing complex narratives.

\bibliographystyle{ACM-Reference-Format}
\balance
\bibliography{ref}

\end{document}